\documentclass{PoS}
\usepackage{amsmath}
\usepackage{slashed}
\usepackage{amsfonts,amsmath,amssymb,bm,graphicx,cite,makeidx,multicol,color}

\title{Neutrino oscillations and evolution in external environments: New effects}

\ShortTitle{Neutrino in external environments: New effects}

\author{\speaker{Alexander Studenikin}\\
        Lomonosov Moscow State University, 119992 Moscow, Russia\\
        Joint Institute for Nuclear Research, 141980 Dubna, Moscow Region, Russia\\
        E-mail: \email{studenik@srd.sinp.msu.ru}}

\abstract{
During a period of about two decades we have realized a programme of
systematic investigations of different aspects of neutrino propagation in extreme
external environments and have predicted and studied several new phenomena
that are engendered by the presence of external magnetic fields and dense matter.
The starting point that underlies the research is the fact that the electromagnetic properties of neutrinos open a window to new physics \cite{Giunti:2014ixa,Studenikin:2008bd, Studenikin:2018vnp}.
In these brief notes, we recall several new phenomena that have been proposed and investigated earlier. In particular, we discuss:
1) the spin light of neutrino in matter, 2) the neutrino energy quantization
in rotating matter, and 3) neutrino start turning mechanism. Then we dwell on
results of recent studies:
4) the effects of interplay of neutrino flavour
and spin oscillations in a magnetic field, 5) the quantum theory of neutrino spin and spin-flavour oscillations engendered by the transversal mater currents, 6) the amplitude modulation of the
flavour neutrino oscillation probability
by the transversal matter current. As for references to the literature,
only those papers are included in which a particular effect was proposed and
considered for the first time and in a number of cases references are also be given
to the most recent articles, which contain detailed references to the available
literature on the issue.}

\FullConference{
European Physical Society Conference on High Energy Physics - EPS-HEP2019 -\\
			10-17 July, 2019\\
			Ghent, Belgium}

\begin{document}


\section{Neutrino spin light and energy quatization in magnetized matter}
{\it Spin light of neutrino.}
The spin light of neutrino ($SL\nu$) is a new mechanism of electromagnetic radiation emitted by a massive neutrino (with a nonzero magnetic moment) moving in external media that was proposed in
\cite{Lobanov-Stud:03} on the basis of the quaziclassical theory of the neutrino spin
evolution in matter.
The quantum theory of the spin light of neutrino in matter  was developed in \cite{Stud-Ternov-PLB:05bar} and it was shown that  $SL\nu$ originates from the neutrino-photon magnetic moment coupling and from the  energy splitting, induced in an external environment, of two different neutrino helicity states characterized by $s= \pm 1$ (see \cite{Stud-Ternov-PLB:05bar,Grig-Stud-Ternov-PLB:05,Lobanov:2005zn,Grigoriev:2017wff, Grigoriev:2012pw} and references therein). The total rate and power of the radiation in homogeneous
neutron matter (far from the threshold),
$  {\Gamma}=2{\mu}^2 G^2_F n^2 p, \ \ \ I={}^2\!\!/_3 {\mu}^2 G^2_F n^2 p^2$,
exhibit strong dependance on the neutrino energy and matter density $n$
($p$ and $\mu$ are the neutrino momentum and magnetic moment). The $SL\nu$ radiation has also a non-trivial polarization properties \cite{Stud-Ternov-PLB:05bar,Grig-Stud-Ternov-PLB:05}. In the nuclear matter, the $SL\nu$ of an ultrarelativistic neutrino is completely circular polarized.

Although this effect is very weak due to smallness of the neutrino magnetic moment,
it can be of interest for astrophysical environments involving compact relativistic
objects because its efficiency is higher, the higher the neutrino energy and
background matter density. In our recent studies \cite{Grigoriev:2017wff, Grigoriev:2012pw} we summarize conditions for best $SL\nu$ efficiency in astrophysical settings and conclude that the most suitable astrophysical site for manifestation of this phenomenon is represented by short
Gamma-Ray Bursts (see, for instance, \cite{Perego-Rosswog-etal-sRGB:2014}) where generation of ultra-high energy neutrinos is anticipated and the matter density can be of the order of the
nuclear density.

{\it Neutrino energy quantization in rotating matter.}
Another interesting phenomenon is the neutrino energy quantization
\cite{Grigoriev:2007zzc} in the case the particle propagation in
a rotating medium. Using the method of exact solutions for the neutrino
wave functions in the external matter \cite{Studenikin:2008qk} it was shown that
the neutrino $\nu_e$ energy spectrum in a rotating and magnetized matter composed
of neutrons is given by
\cite{Grigoriev:2007zzc, Balantsev:2010zw, Studenikin:2012vi, Dvornikov:2015dna}:
$p_0=\sqrt{p_3^2+2N(q_\nu B+2Gn_n\omega)}-Gn_n, \ \ \ G=\frac{G_F}{\sqrt 2}$,
where $N$ is an integer number, $n_n$ is the neutron invariant number density (it is supposed that the direction of the neutrino propagation
coincides with direction of the magnetic field $\bm B$ and the matter rotation
angular velocity $\bm\omega$). The energy spectrum is quantized due to both weak and electromagnetic
interactions (due to possible nonzero neutrino millicharge $q_\nu$) and represents the modified Landau levels of the millicharged neutrino in the rotating magnetized matter. In the quasiclassical treatment neutrino are moving on circular orbits with radius given by
$R=\Omega^{-1}$,
where the effective rotation frequency $\Omega=\omega_m+\omega_c$
is determined \cite{Studenikin:2012vi} by the cyclotron frequency $\omega_c=\frac{q_\nu B}{p_0+Gn_n}$
 and the matter induced frequency $\omega_m=\frac{2Gn_n}{p_0+Gn_n}\omega$.

It is possible to explain \cite{Studenikin:2008qk} the neutrino quasiclassical circular orbits as
a result of action of the
 attractive central force,
${\bf F}_m^{(\nu)}=q^{(\nu)}_m {\bm \beta} \times {\bf B}_m, \ {\bf B}_m= {\bm
\nabla} \times {\bf A}_m, \ {\bf A}_m=n {\bf v}$,
where the effective neutrino ``charge'' in matter (composed of neutrons in the
discussed case) is $q^{(\nu)}_m=-G$, whereas ${\bf B}_m$ and ${\bf A}_m$ play
the roles of effective ``magnetic'' field and the correspondent ``vector
potential''. Like the magnetic part of the Lorentz force, ${\bf F}_m^{(\nu)}$ is
orthogonal to the neutrino speed ${\bm \beta}$. For the most general case
(when the matter density $n$ is not constant) the ``matter
induced Lorentz force'' is given by \cite{Studenikin:2012vi}
${\bf F}_m^{(\nu)}=q^{(\nu)}_m {\bf E}_m + q^{(\nu)}_m {\bm \beta} \times {\bf
B}_m$,
where the effective ``electric'' and  ``magnetic'' fields are
respectively,
${\bf E}_m=-{\bm \nabla}n -{\bf v}\frac{\partial n}{\partial t} -
n\frac{\partial {\bf v}}{\partial t} , \ \ \
{\bf B}_m=n {\bm \nabla} \times {\bf v}-{\bf v} \times {\bm
\nabla}n$, here ${\bf v}$  is the speed of matter.
The
force acting on a neutrino, produced by the first term of the effective
``electric'' field in the neutron matter, was considered in \cite{LoePRL90},
similar quasiclassical treatment of a neutrino motion in the
electron plasma was considered in \cite{OliveiraeSilva:2000ns}.

{\it Neutrino start turning mechanism.}
The effect of the millicharged neutrino energy quantization and the corresponding
neutrino motion on curved orbits in a magnetized matter applied for neutrinos
propagating inside rotating neutron stars provides the best astrophysical bound
for the neutrino millicharge \cite{Studenikin:2012vi}. In the
the magnetized matter the effective Lorentz force discussed in the previous section
disturbs the neutrino trajectories. Obviously, the feedback of
the escaping millicharged neutrinos that are  moving on curved orbits
inside a magnetized  rotating star should effect initial star  rotation.
We have termed this phenomenon the ``Neutrino Star Turning'' ($\nu S T$) mechanism.
In order to avoid  the contradiction  of  $\nu S T$ mechanism impact
with  observational  data  on  pulsars a stringent limit on the neutrino
millichager was obtained \cite{Studenikin:2012vi}: $q_{\nu}<1.3\times10^{-19}e_0$ , that is indeed the best astrophysical upper bound on
$q_{\nu}$.

\section{New effect of neutrino oscillations}
\label{osc_j}
{\it Interplay of neutrino flavour, spin and spin-flavour oscillations
in a constant magnetic field.}
 A new approach  to description of neutrino spin and spin-flavor oscillations in the presence of an arbitrary constant magnetic field has been  developed \cite{Dmitriev:2015ega,Studenikin:2016zdx}
 recently. Within the new approach exact quantum stationary states  are used for classification of neutrino spin states, rather than the neutrino helicity states that are used for this purpose within the customary approach in many published papers. Recall that the helicity states are not stationary in the presence of a magnetic field. It has been shown \cite{Popov:2019nkr}
in particular, that in the presence of the transversal magnetic field $B_{\perp}$ for a given choice of parameters (the  energy and magnetic moments of neutrinos and strength of the magnetic field)  the amplitude of the
flavour oscillations $\nu_e^L \Leftrightarrow \nu_{\mu}^L$ at the vacuum frequency
$\omega_{vac}=\frac{\Delta m^2}{4p}$ is modulated
by the magnetic field frequency $\omega_{B}=\mu B_{\perp}$:
\begin{equation}\label{fl_simp}
P^{(B_{\perp})}_{\nu_{e}^L \rightarrow \nu_{\mu}^L}(t) = \left( 1 - \sin^2(\mu B_{\perp}t) \right) \sin^2 2\theta \sin^2 \frac{\Delta m^2}{4p}t   = \left(1 - P_{\nu_{e}^L \rightarrow \nu_{e}^R}^{cust}\right)
P_{\nu_{e}^L \rightarrow \nu_{\mu}^L}^{cust},
\end{equation}
here $\mu$ is the effective magnetic moment of the electron neutrino and
it is supposed that the following relations between diagonal and transition magnetic
moments in the neutrino mass basis are valid: $\mu_1 = \mu_2, \ \ \mu_{ij}=0, \ i\neq j$.
The customary expression
 $P_{\nu_{e}^L \rightarrow \nu_{\mu}^L}^{cust}(t)=
 \sin^2 2\theta \sin^2 \frac{\Delta m^2}{4p}t$
for the neutrino flavour oscillation probability in vacuum in the presence of the transversal
field $B_{\perp}$ is modified by the factor $1 - P_{\nu_{e}^L \rightarrow \nu_{e}^R}^{cust}$. Since the transition magnetic moment in the flavour basis is absent in the case $\mu_1 = \mu_2$, the process $\nu_e^L \rightarrow \nu_e^R$ is the only way for spin flip, and then $1 - P_{\nu_{e}^L \rightarrow \nu_{e}^R}^{cust}$ should be interpreted as the probability of not changing the neutrino spin polarization. And consequently, this multiplier subtracts the contribution of neutrinos $\nu^{R}_e$ with the opposite polarization
providing the survival of the only contribution from the direct neutrino flavour oscillations $\nu_e^L \Leftrightarrow \nu_{\mu}^L$. Similar results on the important influence of the transversal magnetic field on amplitudes of various types of neutrino oscillations were obtained earlier \cite{Kurashvili:2017zab} on the basis of the exact solution of the effective equation for neutrino evolution in the presence of a magnetic field, which accounts for four neutrino species corresponding to two different flavor states with positive and negative helicities.

Consider the probability of the neutrino spin-flavour oscillations $\nu_e^L \leftrightarrow \nu_{\mu}^R$.
In the case $\mu_1 = \mu_2 = \mu$ we have \cite{Popov:2019nkr}:
\begin{equation}\label{spin_flavour_simplified}
P_{\nu_e^L \rightarrow \nu_{\mu}^R}(t) =  \sin^2(\mu B_{\perp} t)\sin^2 2\theta \sin^2\frac{\Delta m^2}{4p}t.
\end{equation}
The obtained expression (\ref {spin_flavour_simplified}) for the probability can be expressed as a product of two probabilities derived within the customary  two-neutrino-states approach
\begin{equation}\label{PP}
P_{\nu_e^L \rightarrow \nu_{\mu}^R} (t)=
P_{\nu_{e}^L \rightarrow \nu_{\mu}^L}^{cust} (t) P_{\nu_{e}^L \rightarrow \nu_{e}^R}^{cust}(t),
\end{equation}
where $P_{\nu_{e}^L \rightarrow \nu_{\mu}^L}^{cust}(t)=
 \sin^2 2\theta \sin^2 \frac{\Delta m^2}{4p}t$
and
 the usual expression for the neutrino spin oscillation probability
\begin{equation}\label{P_mu_B}
P_{\nu_{e}^L \rightarrow \nu_{e}^R}^{cust}(t) = \sin^2(\mu B_{\perp}t),
\end{equation}
are just the probabilities obtained in the customary approach. A
similar  neutrino spin-flavour oscillations (for the Majorana case) as a
two-step neutrino conversion processes were considered in \cite{Akhmedov:2002mf}
(however, the effect was calculated within perturbation theory
since the probability of neutrino spin-flavour oscillations was supposed to be small).

Finally, in the case of the spin oscillations $\nu_{e}^L \rightarrow \nu_{e}^R$
in the transversal magnetic field $B_{\perp}$ within the the developed approach the
effect of neutrino mixing is accounted for that leads to the modification of the
customary expression $P_{\nu_{e}^L \rightarrow \nu_{e}^R}^{cust}(t)$ for the probability by the factor
$1 - P_{\nu_{e}^{L} \rightarrow \nu_{\mu}^{L}}^{cust}$:
\begin{eqnarray}
\label{spin_simplified}
P_{\nu_{e}^L \rightarrow \nu_{e}^R} = \left[ 1 - \sin^2 2\theta \sin^2\left(\frac{\Delta m^2}{4p}t\right) \right]\sin^2(\mu B_{\perp}t).
\end{eqnarray}

The interplay between different oscillations, that follows from the obtained expressions (\ref{fl_simp}), (\ref{spin_flavour_simplified}) and (\ref{spin_simplified}) for the
oscillation probabilities,  gives rise to interesting phenomena \cite{Popov:2019nkr}:
1) the amplitude modulation of the probability of flavour oscillations $\nu_e^L \rightarrow \nu_{\mu}^L$ in the transversal magnetic field with the magnetic frequency $\omega_{B}=\mu B_{\perp}$ (in the case $\mu_1 = \mu_2$) and more complicated dependence  on the harmonic functions with $\omega_{B}$ for $\mu_1 \neq \mu_2$;
2) the dependence of the spin oscillation probability $P_{\nu_{e}^L \rightarrow \nu_{e}^R}$ on the mass square difference $\Delta m^2$;
3) the appearance of the spin-flavour oscillations in the case $\mu_1=\mu_2$ and $\mu_{12}=0$, the transition goes through the two-step processes $\nu_e^L \rightarrow \nu_{\mu}^L \rightarrow \nu_{\mu}^R$ and $\nu_e^L \rightarrow \nu_{e}^R \rightarrow \nu_{\mu}^R$.

As a result, we predict modifications of the neutrino oscillation patterns that might provide new
 important phenomenological consequences in case of neutrinos propagation in extreme
 astrophysical environments where magnetic fields are present.

{\it Neutrino spin and spin-flavour oscillations
in transversal matter current.}
Another new effect in neutrino oscillations was predicted in \cite{Studenikin:2004bu}:
it was shown the the neutrino spin precession and oscillations can be engendered by
neutrino interactions with transversally moving (or polarized) matter(even in the absence
of a magnetic field). The existence of this effect was also confirmed
in \cite{Cirigliano:2014aoa,Volpe:2015rla,Kartavtsev:2015eva,Dobrynina:2016rwy,Tian:2016hec}.

Initially the effect was predicted \cite{Studenikin:2004bu}
within the semiclassical theory of the neutrino spin evolution in
the transversal matter currents has been recently appended by the direct quantum treatment
of the phenomenon \cite{Studenikin:2016iwq,Pustoshny:2018jxb}.
Consider two flavour neutrinos with two possible helicities
$\nu_{f}= (\nu_{e}^{+}, \nu_{e}^{-}, \nu_{\mu}^{+}, \nu_{\mu}^{-})^T$ in moving matter
composed of neutrons. Each of the flavour neutrinos is a superposition of the neutrino mass states,
  $\nu_{e}^{\pm} =\nu_{1}^{\pm}\cos\theta+\nu_{2}^{\pm}\sin\theta,\
  \nu_{\mu}^{\pm}=-\nu_{1}^{\pm}\sin\theta+\nu_{2}^{\pm}\cos\theta $.
The corresponding neutrino evolution equation is
\begin{equation}\label{schred_eq_fl}
  i\dfrac{d}{dt}\nu_{f}=H^{f}_{v}\nu_{f},
\end{equation}
where the evolution Hamiltonian reads
\begin{equation}\label{H_v}
H^f_v=n\tilde{G}\left( \begin{matrix}
0& (\frac{\eta}{\gamma})_{ee}v_{\perp} &0 &(\frac{\eta}{\gamma})_{e\mu}v_{\perp}\\
(\frac{\eta}{\gamma})_{ee}v_{\perp} & 2(1-v_{\parallel}) &(\frac{\eta}{\gamma})_{e\mu}v_{\perp}&
0\\
0& (\frac{\eta}{\gamma})_{e\mu}v_{\perp} &0 &(\frac{\eta}{\gamma})_{\mu \mu}v_{\perp}\\
(\frac{\eta}{\gamma})_{e\mu}v_{\perp} & 0 &(\frac{\eta}{\gamma})_{\mu\mu}v_{\perp}& 2(1-v_{\parallel})\\
\end{matrix} \right),
\end{equation}
and
\begin{equation}\label{eta}
\Big(\frac{\eta}{\gamma}\Big)_{ee}=
\frac{\cos^2\theta}{\gamma_{11}}+\frac{\sin^2\theta}{\gamma_{22}}, \ \
 \Big(\frac{\eta}{\gamma}\Big)_{\mu\mu}=
\frac{\sin^2\theta}{\gamma_{11}}+\frac{\cos^2\theta}{\gamma_{22}}, \ \
\Big(\frac{\eta}{\gamma}\Big)_{e\mu}=
\frac{\sin 2\theta}{\tilde{\gamma}_{21}},
 \end{equation}
where ${\widetilde{\gamma}_{2 1}}^{-1}=\frac{1}{2}\big(
\gamma_{2}^{-1}-\gamma_{1}^{-1}\big),\ \ \gamma_{\alpha}^{-1}=\frac{m_\alpha}{p^{\nu}_0}$,
$p^{\nu}_0$ is neutrino energy and $\alpha = 1,2 $).

Consider the initial neutrino state
$\nu_{e}^L$ moving in the background with the magnetic field
${\bm B} = {\bm B}_{\parallel}+{\bm B}_{\perp}$ and
nonzero matter current ${\bm j} = {\bm j}_{\parallel}+{\bm j}_{\perp}$.
One of the possible modes of neutrino transitions with the change of helicity is
$\nu_{e}^L\Leftarrow (j_{\perp}, B_{\perp}) \Rightarrow \nu_{e}^R$.
The corresponding
oscillations are governed by the evolution equation \cite{Pustoshny:2018jxb}
\begin{eqnarray}\label{nu_L_nu_R}
	i\frac{d}{dt} \begin{pmatrix}\nu^L_{e} \\ \nu^R_{e} \\  \end{pmatrix}=
	 \left( \begin{matrix}
	(\frac{\mu}{\gamma})_{ee}{B_{||}}+{\widetilde{G}}n(1-{\bm v}{\bm \beta})\\
	 \mu_{ee}B_{\perp}+ (\frac{\eta}{\gamma})_{ee}{\widetilde{G}}nv_{\perp}
	 \end{matrix} \right.
	 \left. \begin{matrix} \mu_{ee}B_{\perp} + (\frac{\eta}{\gamma})_{ee}{\widetilde{G}}nv_{\perp}  \\
	  - (\frac{\mu}{\gamma})_{ee}{B_{||}} -{\widetilde{G}}n(1-{\bm v}{\bm \beta} \end{matrix} \right)
	\begin{pmatrix}\nu^L_{e} \\ \nu^R_{e} \\ \end{pmatrix}.
\end{eqnarray}
Here we constraint our consideration to the binary neutrino transitions and corresponding
oscillations between pairs of the neutrino states.
The oscillation $\nu^L_{e} \Leftarrow (j_{\perp}, B_{\perp}) \Rightarrow \nu^R_{e}$ probability
is given by
\begin{equation}\label{prob_oscillations}
P^{(j_{\perp},B_{\perp})}_{\nu^L_{e} \rightarrow \nu^R_{e}} (x)={E^2_\textmd{eff} \over
{E^{2}_\textmd{eff}+\Delta^{2}_\textmd{eff}}}
\sin^{2}{\pi x \over L_\textmd{eff}}, \ \ \ L_\textmd{eff}={\pi \over \sqrt{E^{2}_\textmd{eff}+\Delta^{2}_\textmd{eff}}},
\end{equation}
where
\begin{equation}\label{E}
E_{eff}= \Big|\mu_{ee}\bm{B}_{\perp} + \Big(\frac{\eta}{\gamma}\Big)
_{ee}{\widetilde{G}}n\bm{v}_{\perp}  \Big|, \ \ \
\Delta_{eff}= \Big|\Big(\frac{\mu}{\gamma}\Big)_{ee}{\bm{B}_{||}}+
{\widetilde{G}}n(1-{\bm v}{\bm \beta}){\bm {\beta}} \Big|.
\end{equation}
From (\ref{prob_oscillations}) and (\ref{E}) it is clearly seen that even in
the absence of the transversal magnetic field
the neutrino spin oscillations $\nu_{e}^L\Leftarrow (j_{\perp}) \Rightarrow \nu_{e}^R$
can be generated by the neutrino interaction with the transversal matter current
$\bm{j}_{\perp}=n\bm{v}_{\perp}$.

In a quite analogous way the transversal matter current can engender the neutrino
spin-flavour conversion and the corresponding oscillations $\nu_{e}^L\Leftarrow (j_{\perp}) \Rightarrow\nu_{\mu}^R$ between two different flavour
states with opposite spin orientations. Note that for apperance of this effect
there is no need for a magnetic field (obviously, the effect is neutrino
magnetic moments independent). For the neutrino evolution $\nu_{e}^L\Leftarrow (j_{\perp}) \Rightarrow\nu_{\mu}^R$ in the case $B=0$
from (\ref{H_v}) we get
\begin{eqnarray}
	i\frac{d}{dt} \begin{pmatrix}\nu^L_{e} \\ \nu^R_{\mu} \\  \end{pmatrix}=
	\left( \begin{matrix}
	 -\Delta M+{\widetilde{G}}n(1-{\bm v}{\bm \beta}) \\
	  (\frac{\eta}{\gamma})_{e\mu}{\widetilde{G}}nv_{\perp} \end{matrix} \right.
	 \left. \begin{matrix}  (\frac{\eta}{\gamma})_{e\mu}{\widetilde{G}}nv_{\perp}  \\
	 \Delta M -{\widetilde{G}}n(1-{\bm v}{\bm \beta})
		\end{matrix} \right)
	\begin{pmatrix}\nu^L_{e} \\ \nu^R_{\mu} \\ \end{pmatrix},
\end{eqnarray}
where $\Delta M=\frac{\Delta m ^2 \cos 2\theta}{4 p^{\nu}_0 }$.
The probability $P_{\nu_{e}^L \rightarrow \nu_{\mu}^R}^{(j_{\perp})}$ of the neutrino spin-flavour
oscillations $\nu^L_{e} \Leftarrow (j_{\perp}) \Rightarrow \nu^R_{\mu}$ is given by the same equation (\ref{prob_oscillations}), but now
\begin{eqnarray}\label{E_1_delta_1}
E_{eff}= \Big|\Big(\frac{\eta}{\gamma}\Big)_{e\mu}{\widetilde{G}}n{v}_{\perp}  \Big|, \ \
\Delta_{eff}
=\Big|\Delta M-{\widetilde{G}}n(1-{\bm v}{\bm \beta})  \Big|.
\end{eqnarray}
From (\ref{E_1_delta_1}) it follows that
the neutrino spin-flavour oscillations $\nu^L_{e} \Leftarrow (j_{\perp}) \Rightarrow \nu^R_{\mu}$
can be generated by the neutrino interaction with the transversal matter current
$\bm{j}_{\perp}=n\bm{v}_{\perp}$.

{\it Amplitude modulation of flavour neutrino oscillation probability
by transversal matter current.}
  From the previous discussions (see \cite{Popov:2019nkr} for details)
  it follows that in the presence of a magnetic field  it is not possible to consider the neutrino flavour and spin oscillations
  as separate phenomena. On the contrary, there is an inherent communication between two.
  In particular, the amplitude of the neutrino flavour oscillations is modulated by the
  magnetic frequency $\omega_{B}=\mu B_{\perp}$.
  The main result of the above discussion
    (see also \cite{Studenikin:2004bu,Pustoshny:2018jxb})
  is the conclusion on the equal role that the transversal magnetic field ${\bm B}_{\perp}$ and the
  transversal matter current $\bm{j}_{\perp}$ plays in generation of the neutrino spin and spin-flavour oscillations.
  From these observations
  \textbf{ we predict a new phenomenon of the modification of
  the flavour neutrino oscillations probability in moving matter under the condition of non-vanishing
  matter transversal current $\bm{j}_{\perp}=n \bm {v}_{\perp}$ }.

  The flavour neutrino oscillation probability
 accounting for this effect can be expressed as follows:
  \begin{equation}\label{flav_mod_current}
P^{(j_{||}+j_{\perp})}_{\nu_{e}^L \rightarrow \nu_{\mu}^L} (t) =  \left(1 - P_{\nu_{e}^L \rightarrow \nu_{e}^R}^{(j_{\perp})} - P_{\nu_{e}^L \rightarrow \nu_{\mu}^R}^{(j_{\perp})}\right)
P_{\nu_{e}^L \rightarrow \nu_{\mu}^L}^{(j_{||})},
\end{equation}
where
$P_{\nu_{e}^L \rightarrow \nu_{\mu}^L}^{(j_{||})}(t) =
\sin^2 2\theta _{eff}\sin^2 \omega_{eff}t$,
is the flavour oscillation probability in moving matter \cite{Grigoriev:2002zr},
 $\omega_{eff}=\frac{\Delta m^2 _{eff}}{4p_{0}^\nu}$,
$\theta _{eff}$
and $\Delta m^2 _{eff}$  are the corresponding quantities modified by the presence of moving matter
(note that in the definition of $\theta _{eff}$
and $\Delta m^2 _{eff}$ only the longitudinal component of matter motion matters).
For the probability of the neutrino spin oscillations  engendered by the transversal current $\bm {j}_{\perp}$
from (\ref{nu_L_nu_R}) we get
\begin{equation}
P^{j_{\perp}}_{\nu_{e}^L \rightarrow \nu_{e}^R}(t)=
\frac{\Big(\frac{\eta}{\gamma}\Big)_{ee}^2 {v}^{2}_{\perp}}
{\Big(\frac{\eta}{\gamma}\Big)_{ee}^2 {v}^{2}_{\perp}+
(1-{\bm v}{\bm \beta})^2}
\sin^2\omega^{j_{\perp}}_{ee}t.
\end{equation}
For the corresponding probability of the neutrino spin-flavour  oscillations due to $\bm {j}_{\perp}$
from (\ref{E_1_delta_1}) we get
\begin{equation}
P^{j_{\perp}}_{\nu_{e}^L \rightarrow \nu_{\mu}^R}(t)=
\frac{\Big(\frac{\eta}{\gamma}\Big)_{e\mu}^2 {v}^{2}_{\perp}}
{\Big(\frac{\eta}{\gamma}\Big)_{e\mu}^2 {v}^{2}_{\perp}+
\Big(\frac{\Delta M}{{\widetilde{G}}n} - (1-{\bm v}
{\bm \beta})\Big)^2}
\sin^2\omega^{j_{\perp}}_{e\mu}t.
\end{equation}
The discussed new effect of the modification of the
flavour oscillations $\nu_{e}^L\Leftarrow (j_{||}, j_{\perp})\Rightarrow \nu_{\mu}^L$ probability
is the result of an interplay of oscillations on a customary
flavour oscillation frequency in moving matter $\omega_{eff}$ and
two additional oscillations with changing the neutrino polarization,
the neutrino spin $\nu_{e}^L\Leftarrow (j_{\perp}) \Rightarrow \nu_{e}^R$ and spin-flavour
$\nu_{e}^L\Leftarrow (j_{\perp}) \Rightarrow \nu_{\mu}^R$ oscilations, that are
governed by two
characteristic frequencies:
\begin{equation}\label{omega_1}
  \omega^{j_{\perp}}_{ee}={\widetilde{G}}n{\sqrt{\Big(\frac{\eta}{\gamma}\Big)_{ee}^2 {v}^{2}_{\perp} +
  (1-{\bm v}{\bm \beta})^2}} ,\ \ \
  \omega^{j_{\perp}}_{e\mu}={\widetilde{G}}n
  {\sqrt{\Big(\frac{\eta}{\gamma}\Big)_{e\mu}^2 {v}^{2}_{\perp}
  +
  \Big(\frac{\Delta M}{{\widetilde{G}}n}-(1-{\bm v}{\bm \beta})\Big)^2}}.
\end{equation}
The discussed interplay of oscillations on different frequencies should be accounted for in analysis
of propagation of neutrino fluxes in astrophysical environments.

\end{document}